\documentclass[12pt,tightenlines]{revtex4}
\usepackage{epsfig,equations}

\newenvironment{equation*}{\begin{displaymath}}{\end{displaymath}}

\newcommand{\Scri}{\mbox{$\cal J$}}

\newcommand{\DIII}{\,{}^{\scriptscriptstyle(3)\!\!\!\:}\nabla}

\newcommand{\RIII}{\,{}^{\scriptscriptstyle(3)\!\!\!\:}R}
\newcommand{\ROI}{\,{}^{\scriptscriptstyle(0,1)\!\!\!\:}\hat R}
\newcommand{\RII}{\,{}^{\scriptscriptstyle(1,1)\!\!\!\:}\hat R}

\def\@warning#1{\typeout{LaTeX Warning [l.\the\inputlineno]: #1.}}

\begin{document}
\title{On the Effect of Constraint Enforcement on the Quality of
  Numerical Solutions in General Relativity}
\author{Florian Siebel}
\affiliation{Max-Planck-Institut f\"ur Astrophysik\\
Karl-Schwarzschild-Str. 1\\
85741 Garching, Germany\\
florian@mpa-garching.mpg.de\\}
\author{Peter H\"ubner}
\affiliation{Linienstra\ss{}e 45a\\
82041 Oberhaching, Germany\\
pth@epost.de\\}
\begin{abstract}
  In Brodbeck et al 1999~\cite{BFHR} it has been shown that the linearised
  time evolution equations of general relativity can be extended to 
  a system whose solutions asymptotically approach solutions of the 
  constraints.
  \\
  In this paper we extend the non-linear equations in similar
  ways and investigate the effect of various possibilities by
  numerical means.
  Although we were not able to make the constraint submanifold an
  attractor for all solutions of the extended system, we were able to
  significantly reduce the growth of the numerical violation of
  the constraints.
  Contrary to our expectations this improvement did not imply a
  numerical solution closer to the exact solution, and therefore did
  not improve the quality of the numerical solution.
\end{abstract}
\maketitle
\section{Introduction}
\label{EINS}
Many physical theories are based upon systems of partial differential
equations which contain more equations than variables like Maxwell's
equations or general relativity.
The initial data for the time evolution equations cannot be given freely, 
they must satisfy constraints.
It is necessary for the consistency of the theory, that for any data
of the time evolution equations which initially satisfy the constraints, 
the constraints are satisfied for all times.
This property is called ``propagation of constraints''.
\\
Let us consider Maxwell's equations in vacuum as a simple example.
The time evolution equations tell us that the time derivative of the
electric and magnetic field are proportional to the curl of the
magnetic and electric field.
The vanishing of the divergence of the electric and magnetic field are
the constraints.
It can easily be shown that the constraints propagate.
\\
In cases where the solutions of a system of partial differential
equations are determined by numerical means we cannot expect to get an
exact propagation of the constraints.
Due to the discretisation of the equations the numerical solution
deviates from the exact solution by the discretisation error.
As a consequence, the constraints are not fulfilled exactly after
having evolved for some time, even if the initial data solved the
constraints.
In the spirit of~\cite{Cho} we call a discretisation of the time
evolution equations compatible with the constraints, if the numerical
violation of the constraints has the same convergence order as the
discretisation of the time evolution equations.
Unfortunately the experience of numerical relativity shows that
compatibility is not sufficient for obtaining numerical solutions with
small numerical violations of the constraints.
In many cases the violation of the constraints seems to grow at least
exponentially with time.
This effect is believed to be a major contribution to the numerical
error of numerically calculated solutions.

In this work we examine the effect of changing the evolution equations
outside the submanifold of data on which the constraints are
satisfied.
Although our method is different, we should mention that already
in~\cite{Det} such a change has been suggested.
Furthermore, to our knowledge, we perform the first systematic
analysis of the correlation between the violation of the constraints
and the quality of numerical solutions in general relativity.
\\
As the solutions of the field equations of general relativity 
satisfy the constraint equations for all times, the solutions are not
affected by modifications of the evolution equations for data which do
not satisfy the constraints.
Let us denote the subspace of the function space of
solutions to the evolution equations which satisfy the constraints as
``constraint submanifold''.
In~\cite{BFHR} it has been proven, that at least for the linearised
Einstein equations the constraint submanifold can be made an attractor
for the linearised time evolution equations.
\\
If the solution of the evolution equations automatically approaches
the constraint submanifold, the system of evolution equations carries
a dissipative term in it, and therefore, the numerical solution will
also approach the constraint submanifold provided the grid is not too
coarse.
Therefore, to avoid a numerical violation of the constraints, it is
sufficient to make the constraint submanifold an attractor of the
(modified) evolution equations.
In such a case the constraint submanifold is `asymptotically stable'.  

In Brodbeck et al.~\cite{BFHR} a general method has been proposed to
derive symmetric hyperbolic extensions of symmetric hyperbolic
evolution equations with first order constraints which are promising
candidates for asymptotic stability.
These extended systems are called $\lambda$-systems.
In the same article it has also been proven, that at least in the case
of the linearised Einstein equations there exist parameters such that
the constraint submanifold is indeed an attractor for the modified
evolution equations.
\\
As the extension of the analysis to the non-linear Einstein equations
seemed to be beyond the scope of present analytical techniques, we
took a numerical approach in this paper and investigated the following
questions:
Firstly, can we, simply by way of numerical experiments, find a
$\lambda$-system for the non-linear Einstein equations for which the
constraint submanifold is attractive?
And secondly, is the numerical solution of the modified systems closer
to the exact solution than the numerical solution of the unmodified
system?
To reduce the complexity and to have exact solutions available to
compare with, we have restricted our investigations to solutions with
two Killing vectors.
\\
In our experiments we were able to find a variety of $\lambda$-systems
for which the violation of all constraints is improved. 
However, we did not find a single system for which the constraint
submanifold is asymptotically stable.
Surprisingly, the improvement in the constraint violation did not
imply an improvement of the numerical solution. 
\\
It is important to mention that a 
general attractive force towards the constraint submanifold does 
not guarantee the numerical solution to approach the exact solution 
corresponding to the initial data used. Regardless of the system of 
the field equations of general relativity, there are additional degrees 
of freedom which can be affected by the additional terms in the 
$\lambda$-system. In our numerical experiments, these additional degrees 
of freedom were affected in such a way that in general the numerical 
solution of the modified system was not closer to the exact solution, 
even if it was closer to the constraint submanifold.

This paper is structured as follows:
In chapter~\ref{ZWEI} we introduce parametrised $\lambda$-systems and
describe the simplifications implied by the symmetry assumptions.
In the next chapter we sketch the numerical implementation, recall
important features of the exact solutions used in the comparisons, and 
define the measures used for quality assessments.
Chapter~\ref{VIER} contains the actual numerical investigations, where
we describe the performed probing of the parameter space. 
Using selected examples we demonstrate the influence of the individual 
parameters on the quality of the numerical solution.
\section{The parametrisation of the $\lambda$-system}
\label{ZWEI}
The construction of a $\lambda$-system is based on a split of the
system of equations into symmetric hyperbolic evolution equations and
first order constraints~\cite{BFHR}.
There are various possibilities to write Einstein's equations in a 
form like this.
\\
In our work, we take the conformal field equations~\cite{Fri,Fri1,Fri2} in the
first-order formulation described in~\cite{Hue}.
Looking at the equations (13) and (14) of~\cite{Hue}, it is easy to
see that Einstein's equations and their extension, the conformal
field equations, are a ``quasilinear version'' of Maxwell's
equations.
We use the conformal field equations instead of Einstein's
equations to obtain an easy and well defined treatment of grid
boundaries, as discussed in~\cite{Hue}.
Since we are primarily interested in the effect of the
non-linearities, we can reduce the computational complexity by
restricting ourselves to asymptotically A3 spacetimes~\cite{Hue2},
which are spacetimes with two commuting, hypersurface orthogonal
Killing vector fields.
We align the $y$ and $z$ coordinates with the Killing orbits. 
Our solutions do therefore not depend on the spacelike coordinates $y$ and
$z$. Under these symmetry assumptions, the conformal field equations can be
written in the following form
\begin{subequations}
\label{Einstein}
\begin{eqnarray}
\label{Einstein1}
\tilde{\mathbf{A}} \frac{\partial}{\partial t} g + {\mathbf{A}}
\frac{\partial}{ \partial x} g - b_{g} & = & 0\\ 
\label{Einstein2}
\frac{\partial}{\partial t} f  - b_{f} & = & 0\\
\label{Einstein3}
\frac{\partial}{\partial x} f - c_{f} & = & 0,
\end{eqnarray}
\end{subequations}
with a time coordinate $t$ labelling the spacelike hypersurfaces,
$x$ being the non-Killing spacelike coordinate, and
\begin{subequations}
\begin{eqnarray}
 g & = & (k_{11}, \gamma^{1}{}_{11}, E_{22}, E_{33}, B_{23}, \ROI_{1}, \RII_{11})\\
 f & = & (h_{11}, h_{22}, h_{33}, k_{22}, k_{33}, \gamma^{1}{}_{22},
 \gamma^{1}{}_{33}, E_{11}, \RII_{22}, \RII_{33}, \Omega, \Omega_{0},
 \Omega_{1}, \omega).
\end{eqnarray}
\end{subequations}
The tensor $h_{ab}$ is the 3-metric, $k_{ab}$ the extrinsic
curvature of the  spacelike hypersurfaces, $\gamma^{a}{}_{bc}$
the connection for $h_{ab}$, $\ROI_{a}$ and $\RII_{ab}$ parts of
the tracefree part of the Ricci tensor, $E_{ab}$ and $B_{ab}$ the
electric and magnetic part of the conformal Weyl tensor, $\Omega$ the
conformal factor, $\Omega_0$ and $\Omega_1$ its normal and space
derivative, and $\omega$ a second derivative of the conformal factor,
as described in more detail in~\cite{Hue}.
The symmetric matrices $\tilde{\mathbf{A}}$, $\mathbf{A}$ and the vectors
$b_{g}$, $b_{f}$ and $c_{f}$ depend on $g$, $f$ and gauge functions.
The matrix $\tilde{\mathbf{A}}$ is positive definite, hence the
system consisting of the equations (\ref{Einstein1}) and
(\ref{Einstein2}) is symmetric hyperbolic.
\\
The variables, which are functions of $t$ and $x$ only, have
been split into two classes, called $g$ and $f$.
For the variables denoted by $f$ the system contains evolution
equations and constraints.
Since there are only evolution equations for the variables $g$, these
represent the degrees of freedom.

When we evolve initial data forward in time by means of the evolution
equations (\ref{Einstein1}) and (\ref{Einstein2}), the
constraints~(\ref{Einstein3}) fulfil an evolution equation from which
the propagation of the constraints can be derived.
\\
In the following we will call (\ref{Einstein1}) and (\ref{Einstein2})
the `unextended system'. 
To obtain the new, extended system, the $\lambda$-system, additional
variables, called $\lambda$, are introduced and the constraint equations
are extended to evolution equations for the new variables: 
\begin{subequations}
\label{lambda}
\begin{eqnarray}
  \tilde{\mathbf{A}} \frac{\partial}{\partial t} g  
  + \ \ \ {\mathbf{A}}  \frac{\partial}{ \partial x} g  
  - \ \  \ b_{g}  +  {\mathbf{C}} \lambda & = & 0
\\
  \frac{\partial}{\partial t} f  + \ \ \ {\mathbf{B}}
  \frac{\partial}{\partial x} \lambda  - \ \ \ b_{f}  +  {\mathbf{D}}
  \lambda & = & 0
\\ 
\label{lambda3}
  \frac{\partial}{\partial t} \lambda  + \ {\mathbf{B}}^{T}
  \frac{\partial}{\partial x} f  - {\mathbf{B}}^{T} c_{f}  +
  {\mathbf{E}} \lambda & = & 0. 
\end{eqnarray}
\end{subequations}
The quantities $\mathbf{B}, \mathbf{C}, \mathbf{D}$, and $\mathbf{E}$
are matrices.
$\mathbf{B}^{T}$ denotes the transposed matrix of $\mathbf{B}$. 
This system is constructed in such a way that 
\begin{enumerate}
\item it is symmetric hyperbolic.
\item in the case in which the variables $\lambda$ vanish identically
  the system is reduced to the original system~(\ref{Einstein}). 
\end{enumerate} 
Due to the second requirement, the $\lambda$-system is a
generalisation of the original system.
The first requirement implies well-posedness of the initial value problem.  
Apart from the restrictions resulting from the two conditions above, the
choice of the parameters $\mathbf{B}, \mathbf{C}, \mathbf{D}$, and
$\mathbf{E}$ is free.
It is the aim to choose them in such a way that for all
solutions of the system the variables $\lambda$ decay, which then
implies that the solution is driven towards a solution of the
constraints.
\\ 
We will now shortly explain why we have introduced these parameters.
Let us assume that the constraint equations are not fulfilled
exactly, i.e.~$\partial_x f - c_{f} \neq 0$, and $\lambda=0$ initially.
In the case of vanishing $\mathbf{E}$, the variables $\lambda$ are the 
time integral of the violation of constraints - as a result of the new 
evolution equation~(\ref{lambda3}).
For  non-vanishing $\mathbf{E}$, in addition, the $\mathbf{E}
\lambda$-term will have a damping or amplifying effect onto the
variables $\lambda$, depending on the eigenvalues of $\mathbf{E}$.
For positive eigenvalues we expect damping, for negative eigenvalues
amplification.
The information about the violation of constraints, saved in the
variables $\lambda$, is coupled back to the variables $(g,f)$ by the  
terms $\mathbf{B} \partial_x \lambda$, $\mathbf{D} \lambda$, and
$\mathbf{C} \lambda$.
\\
As the conformal field equations are a generalisation of Einstein's 
equations we can relate the constraints~(\ref{Einstein3}), or more 
general, the constraints of the conformal field equations without 
symmetries~\cite{Hue}, to the momentum and Hamiltonian constraints of 
the standard 3+1 equations. The Hamiltonian constraint and 
the momentum constraint form a 
subset of the constraints used in our system. Contracting equation (14b)
of~\cite{Hue} with the 3-metric $h^{bc}$ and restricting it to the case
of $\Omega = 1$, we recover the momentum constraint 
$\DIII_{b} (k^{ab}-h^{ab}k)= 0$, with $k = k_{ab} h^{ab}$. 
Similarly, contracting equation (14c) of 
~\cite{Hue} twice and evaluating the 3-Christoffel symbols, 
one can deduce the Hamiltonian constraint
$\RIII + k^{2} - k_{ab} k^{ab} = 0$, where $\RIII$ denotes the Ricci 
scalar of the 3-metric. 
\section{The numerical implementation}
\label{DREI}
In this section we will describe the basic elements of the numerical
implementation which we use to compare the quality of solutions
obtained by solving the $\lambda$-system with the quality
of solutions obtained by solving the unextended system.
These elements are the construction of initial data, the scheme to 
numerically integrate the time evolution equations, and the measures 
used to assess the quality of our numerical solutions.
In addition we will briefly describe the exact solutions which we used as
reference solutions.
\subsection{Constructing hyperboloidal initial data}
In order to analyse the numerical behaviour of the $\lambda$-system,
we first have to construct initial data for the conformal field
equations.
These data are called hyperboloidal initial data.
\\
In~\cite[section~2]{Hue3} the procedure of calculating initial data
has been described in detail for the case without any symmetry assumptions.
We slightly modified the procedure by making use of the symmetry
assumptions, namely that our spatial grid is only
one-dimensional~(1D).
For an exact solution we prescribe the 4-metric $g_{ab}$ and the
conformal factor $\Omega$ as functions of $(t,x)$.
From those we calculate our variables $(g,f)$ and the gauge source
functions numerically.
The code has also got the functionality to perform a coordinate transformation
to express the exact solution in new coordinates $(t',x')$.
\\
In the calculations presented in this paper we used $\lambda=0$ as
initial setting for $\lambda$. 
\subsection{The integration of the time evolution equations}
In order to discretise the evolution equation
\begin{equation}
\label{evolution}
  \partial_{t} \mathbf{u} + \underline{\underline{A}}(\mathbf{u}) 
  \ \partial_{x} \mathbf{u} = \underline{b}(\mathbf{u})
\end{equation}
for the vector of variables $\mathbf{u}$, we adjust the second-order
scheme described in~\cite[section~3.1]{Hue3} to our symmetry
assumption, i.e.\  we perform a Strang splitting ansatz to split
equation~(\ref{evolution}) into a principal part
\begin{equation}
\label{principal}
  \partial_t \mathbf{u} + 
  \underline{\underline{A}}(\mathbf{u}) \ \partial_{x} \mathbf{u} = 0
\end{equation}
and a source part
\begin{equation}
\label{source}
  \partial_t \mathbf{u} = \underline{b}(\mathbf{u}).
\end{equation}
We then solve the principal part~(\ref{principal}) by the rotated Richtmyer
scheme. 
With one spatial dimension, this is equivalent to the standard second-order
Lax-Wendroff method~\cite{LWe}.
The source part is integrated by the pseudo-implicit Heun-scheme. 
As described in~\cite{Hue3}, principal and source part are combined
in different order, depending on whether the time step is odd or even,
to achieve second-order convergence.

In~\cite{Hue3} it was shown how superior a 4th order
scheme would be.
In a normal application this superiority would be a big
advantage --- here it is a disadvantage, however:
We are going to analyse the impact of a drift away from the constraint
submanifold on the quality of the solution.
This drift originates in the discretisation error.
If the scheme is very accurate, the drift is very small, too.
But then, the differences in the quality of the numerical solutions
are also very small, which makes it harder to distinguish between them.
This explains why we use the second-order scheme.

In the runs described in part~\ref{VIER} we cover the spacelike hypersurface with 161 grid points.
The length of the time step is chosen dynamically by evaluating the
Courant-Friedrichs-Lewy condition for each time slice.
If not explicitly stated otherwise, we use half of what would be
allowed by the Courant-Friedrichs-Lewy condition.
\subsection{Measures of quality}
We use the following measures to
analyse the quality of a numerical solution.
\\
Firstly, we determine the numerical violation of the constraints as a
measure of the distance from the constraint submanifold.
To be able to present our findings with a limited number of plots we
condense the information by using the ``norm'' 
\begin{equation}
\label{normconstr}
  ||\Delta_{{\cal C}}||(t) := 
  \sqrt{\sum_{l}^{} \int_{\Omega > 0} {\cal C}_{l}(t,x)^{2} dx},
\end{equation}
where the summation includes all constraints ${\cal C}_{l}(t,x)$.
In the actual numerical calculations the integral is of course
replaced by a sum over all grid points with $\Omega>0$,
which represent physical spacetime.
In the conformal approach to numerical relativity grid points with
$\Omega<0$ represent a formal extension of the grid without
physical relevance.
Therefore, it would be wrong to include them into the measure.
\\
Secondly, we compare our numerical solution to the exact
solution.
Since the numerical calculation of $(\RII_{ab},E_{ab},B_{ab})$ from
the given solution $(g_{ab},\Omega)$ involves solving elliptic
equations (cf.~\cite{Hue3}), we compare $(\Omega \RII_{ab},\Omega^2
E_{ab},\Omega B_{ab})$ to the corresponding quantities from the
exact solution and call the result ``pseudodifference'' ${\cal P}_{l}(t,x)$.
In~\cite{Hue3} it was found that with respect to the relative error
this is equivalent to the difference in the variables.
Again, to be able to present our findings with a limited number of
plots, we condense the information by using the ``norm''
\begin{equation}
\label{normpseudo}
  ||\Delta_{{\cal P}}||(t) := 
  \sqrt{\sum_{l}^{} \int_{\Omega > 0} {\cal P}_{l}(t,x)^{2} dx},
\end{equation}
where the summation includes the variables in the tuple $(g,f)$, but
does not include the variables $\lambda$.
\subsection{The exact solutions used in the numerical experiments}
\begin{figure}[htbp]
  \begin{center}
\begin{picture}(0,0)%
\epsfig{file=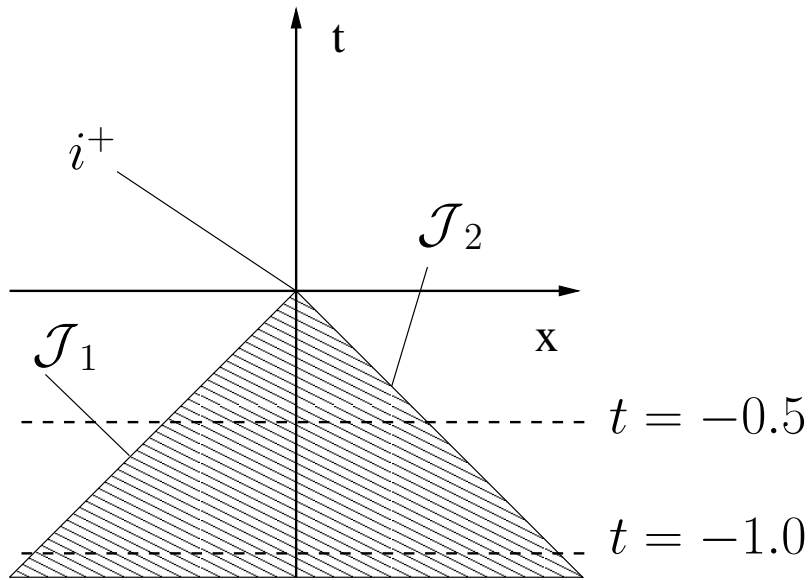}%
\end{picture}%
\setlength{\unitlength}{0.00083300in}%
\begingroup\makeatletter\ifx\SetFigFont\undefined%
\gdef\SetFigFont#1#2#3#4#5{%
  \reset@font\fontsize{#1}{#2pt}%
  \fontfamily{#3}\fontseries{#4}\fontshape{#5}%
  \selectfont}%
\fi\endgroup%
\begin{picture}(3772,3644)(1629,-5858)
\put(1801,-4486){\makebox(0,0)[lb]{\smash{\SetFigFont{20}{24.0}{\familydefault}{\mddefault}{\updefault}$\Scri_1$}}}
\put(5401,-5686){\makebox(0,0)[lb]{\smash{\SetFigFont{20}{24.0}{\familydefault}{\mddefault}{\updefault}$t=-1.0$}}}
\put(2026,-3286){\makebox(0,0)[lb]{\smash{\SetFigFont{20}{24.0}{\familydefault}{\mddefault}{\updefault}$i^{+}$}}}
\put(5401,-4936){\makebox(0,0)[lb]{\smash{\SetFigFont{20}{24.0}{\familydefault}{\mddefault}{\updefault}$t=-0.5$}}}
\put(4201,-3736){\makebox(0,0)[lb]{\smash{\SetFigFont{20}{24.0}{\familydefault}{\mddefault}{\updefault}$\Scri_2$}}}
\end{picture}
    \caption{Conformal diagram of an asymptotically A3 solution. The shaded region corresponds to the physical spacetime.}
    \label{plot1}
  \end{center}
\end{figure}

Only a few exact solutions of Einstein's vacuum equations possess
both, high symmetries and time dependence, or even better
gravitational radiation.
The asymptotically A3 solutions do.
They can be written as
\begin{subequations}
\label{A3}
\begin{eqnarray}
\label{a3}
  g & = & \frac {4 \sqrt{2}}{\sqrt{t^{2}+x^{2}}} e^{M}(-dt^{2}+ dx^{2}) 
          + \frac {1}{2}(t^{2}+x^{2}) (e^{W} dy^{2}+e^{-W}dz^{2})
\\
  \Omega & = & \frac{1}{4}(t^{2}-x^{2}),
\end{eqnarray}
\end{subequations}
where the functions $M(t,x)$ and $W(t,x)$ are solutions of certain
differential equations (cf.~\cite{Hue2}).
\\
We restricted our analysis to two cases, $M \equiv 0$ and $W \equiv 0$,
the A3 solution on the one hand, and $M =
-\frac{1}{256}(t^{2}+x^{2})^{2}$ and $W=\frac{1}{8}(t^{2}-x^{2})$ on
the other hand.
The second case, unlike the first one, contains gravitational radiation.
Both solutions behaved very similarly in our numerical experiments.
The amount of gravitational wave content does not seem to be
significant for the drift away from the constraint submanifold.
For shortness we only present calculations done with the A3 solution.
\\
The solutions given in (\ref{A3}) are already extended beyond the two
null-infinities $\Scri_1$ and $\Scri_2$ at $t = -x$ and $t = +x$,
the shaded region in FIG.~\ref{plot1} shows the physical part.
In this representation the point $(0,0)$ represents future time-like
infinity $i^{+}$. 
The metric and curvature quantities diverge there. 
When approaching this point in our numerical evolution, the absolute value
of several quantities must increase strongly, which implies increasing
absolute errors.
However, we confirmed that, even when approaching $i^{+}$, our
numerical scheme is second-order convergent, the difference between the
numerical and the exact solution as well as the violation of constraints converges with a convergence rate of two.
In the numerical experiments we start our calculation at $t=-1$ and stop at
$t=-1/2$.
\\
In other coordinates FIG.~\ref{plot1} looks different, e.g. one can choose the
coordinates such that the \Scri{}s are at constant $x'$ values and that
$i^{+}$ lies at a conformal time $t'=\infty$.
Since our findings are not influenced by this change of coordinates we
refrain from a presentation.
\section{The quality of the numerical solution: parameter study}
\label{VIER}
In this section we discuss the effect of the various parameters in the
$\lambda$-system (\ref{lambda}).
In the case of the 1D conformal field equations, the matrices
$\mathbf{B}, \mathbf{C}, \mathbf{D}$ and $\mathbf{E}$ take values in
${\mathbf{R}}^{14,14}$, ${\mathbf{R}}^{7,14}$, ${\mathbf{R}}^{14,14}$,
and ${\mathbf{R}}^{14,14}$. 
Since the parameter space is infinite, we had to restrict ourselves to
exemplary cases, an exhaustive study by numerical means is impossible.
\subsection{Overview}
\begin{figure}[htpb]
\noindent
\begin{minipage}[t]{.46\linewidth}
 \centering\epsfig{figure=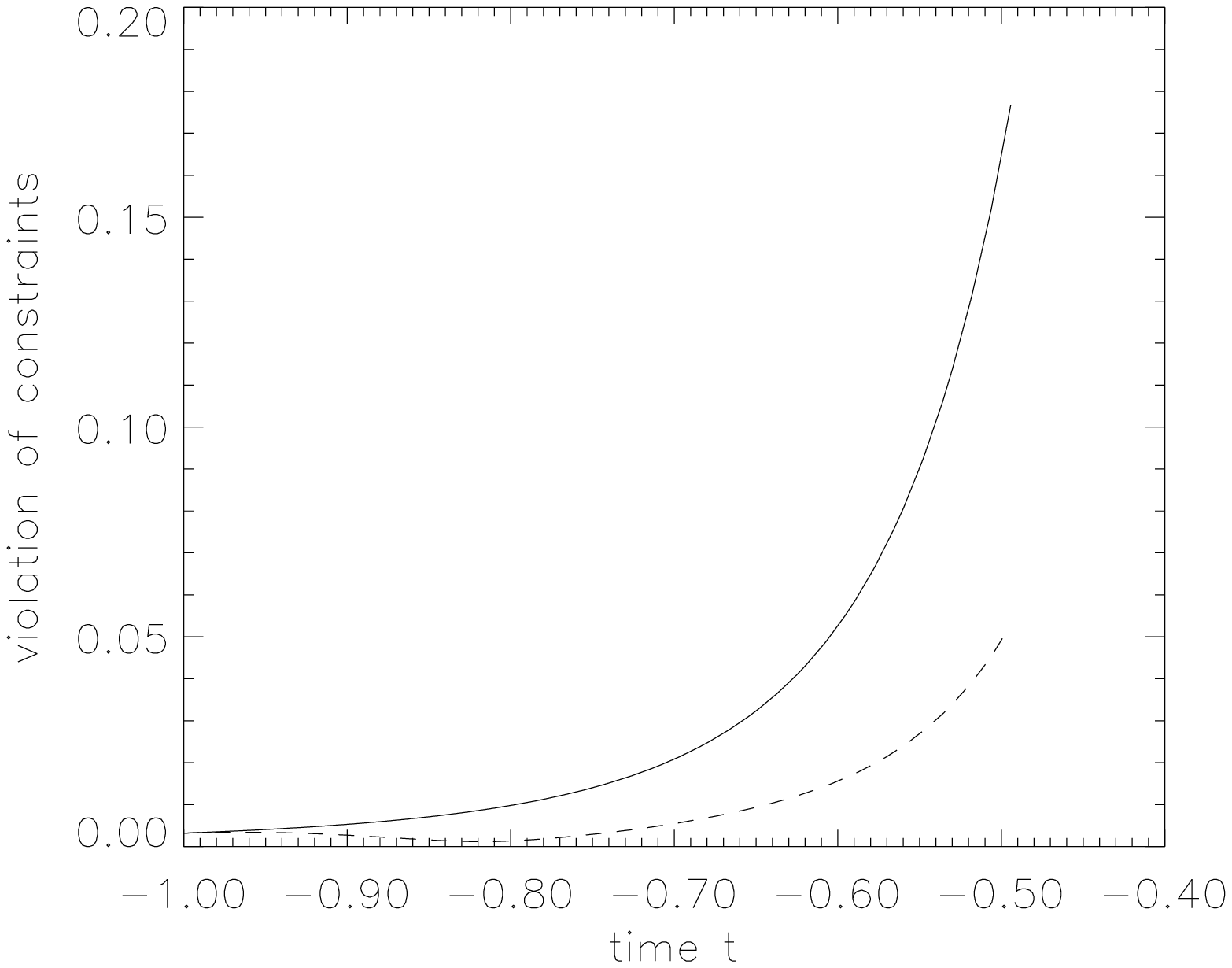,width=\linewidth}
  \caption{Constraint norm $||\Delta_{{\cal C}}||(t)$ for the 
    $\lambda$-system of run~7 (dashed line) in comparison to the unextended
    system (solid line).}
 \label{plot2}
\end{minipage}\hfill
\begin{minipage}[t]{.46\linewidth}
 \centering\epsfig{figure=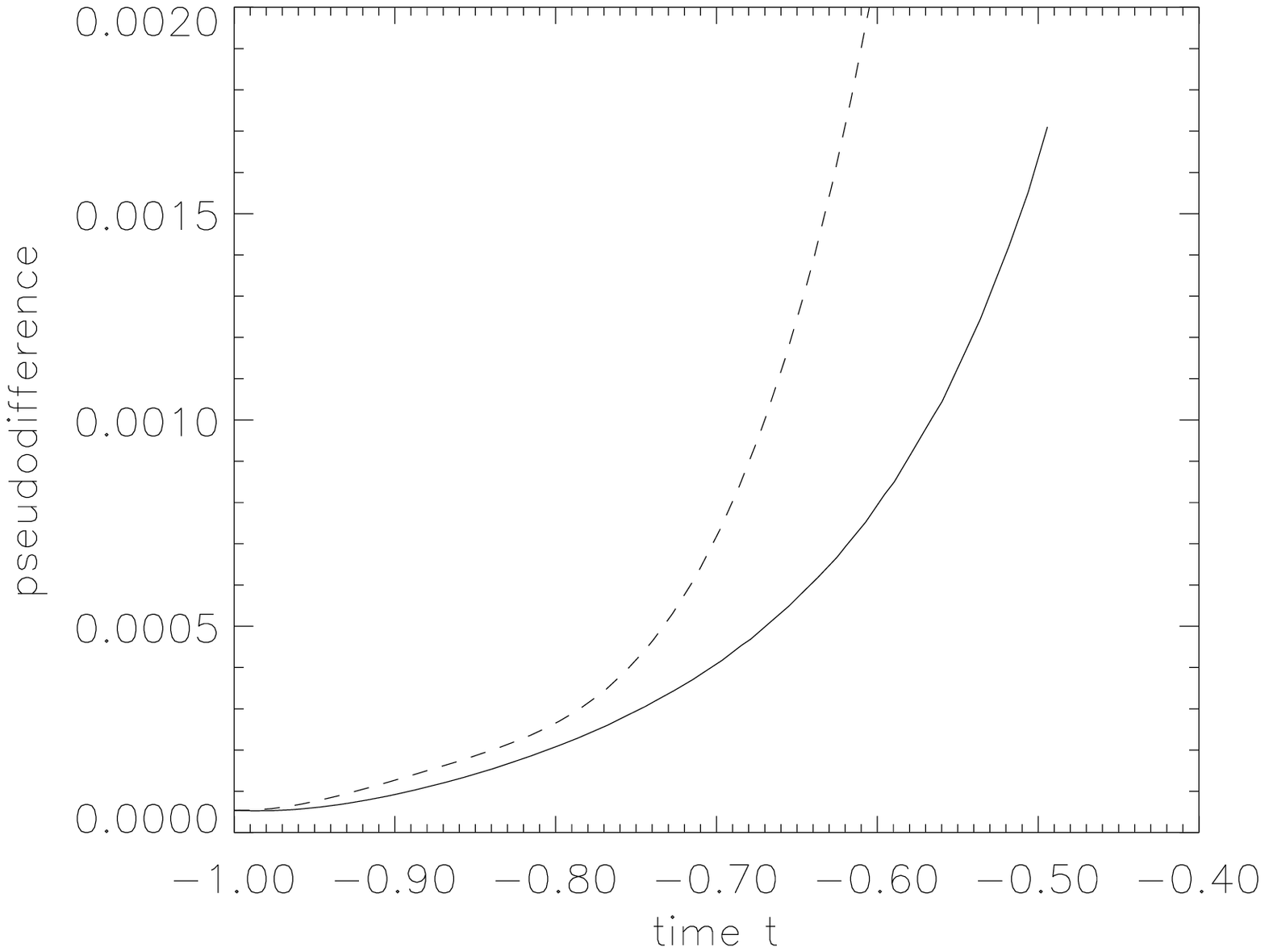,width=\linewidth}
 \caption{Pseudodifference norm $||\Delta_{{\cal P}}||(t)$ 
   for the $\lambda$-system of run~7 (dashed line) in comparison to the
   unextended system (solid line).}
 \label{plot3}
\end{minipage}
\end{figure}

Before going into detail, we discuss the main result of our
numerical experiments. 
Although we were not able to find a suitable choice of parameters such that
the constraint submanifold became an attractor for all times, we were
able to improve the violation of constraints up to a factor 5 compared
to a numerical evolution of the unextended system.
There are strong numerical indications that this improvement is
obtained at the cost of a solution which is worse with respect to
the pseudodifference norm. 
As an example, the dashed lines in FIG.~\ref{plot2} and~\ref{plot3}
show the norms for the violation of constraints~(\ref{normconstr}) and
the pseudodifference~(\ref{normpseudo}) for a choice of parameters
${\mathbf{B}}=3 \cdot \underline{\underline{1}}$, ${\mathbf{C}} = 0$,
${\mathbf{D}}= 0$, and ${{\mathbf{E}}} = \underline{\underline{1}}$ in
the $\lambda$-system (by $\underline{\underline{1}}$ we denote the $14
\times 14$ unit matrix).
The solid lines show the corresponding results for an evolution of the
unextended system~(\ref{Einstein}).
FIG.~\ref{plot2} shows that the $\lambda$-system initially attracts
towards the constraint submanifold (until $t = -0.85$), before the
violation increases with time.
FIG.~\ref{plot3} shows that the pseudodifference norm at the same time
increases more rapidly right from the beginning than in the numerical
solution of the unextended system.

In TABLE 1, we present a summary of the performed numerical
experiments.
\begin{table}[htpb]
\begin{center}
\begin{minipage}{0.8\linewidth}
\begin{center}
\footnotesize
\begin{tabular}[t]{c||c|c|c|c||c|c||}
no of& \multicolumn{4}{c||}{parameter} & \multicolumn{2}{c||}{norm} \\
run& \multicolumn{1}{c|}{$\mathbf{B}$} & ${\mathbf{C}}^{\alpha)}$ &
$\mathbf{D}$ & $\mathbf{E}$ & $||\Delta_{{\cal C}}||$ & $||\Delta_{{\cal P}}||$ \\ \hline \hline
1 & $1$ & $(1,0)$ & $1$ & $1$ & $-$ & $\uparrow$ \\ \hline
2 & $1$ & $(1,0)$ & $1_{\lambda^{2}}{}^{\beta)}$ &
$1_{\lambda^{2}}{}^{\beta)}$ & $-$ 
& $ \uparrow$ \\ \hline
3 & $1$ & $(1,0)$ & $1$ & $1_{\lambda^{2}}{}^{\beta)}$ & $-$ & $ \uparrow$\\ \hline
4 & $1$ & $(1,0)$ & $1_{\lambda^{2}}{}^{\beta)}$ & $1$ & $-$ & $
\uparrow$\\ \hline \hline
5 & $|b| \in ]0,1]$ & $(0,0)$ & $0$ & $1$ & $-$ & $ \uparrow$\\ \hline
6 & $1.2$ & $(0,0)$ & $0$ & $1$ & $ - $ & $ \uparrow$\\ \hline
7 & $3$ & $(0,0)$ & $0$ & $1$ & $-$ & $+$ \\ \hline \hline
8 & $1$ & $(0,0)$ & $0$ & $10$ &$ + $ & $+$\\ \hline
9 &$1$ & $(0,0)$ & $0$ & $e \in [-1,1]$ &$-$ & $\uparrow$\\ \hline
10 &$1$ & $(0,0)$ & $0$ & $-3$ &$-$ & $-$\\ \hline
11 & $1$ & $(0,0)$ & $0$ & $-10$ & $ \uparrow $ & $+$\\ \hline
12 & $1$ & $(0,0)$ & $0$ & $({\mathbf{E}})_{11}=({\mathbf{E}})_{88}=-3^{\gamma)}$ & $-$ & $-$ \\ \hline 
13 & $1$ & $(0,0)$ & $0$ & $({\mathbf{E}})_{11}=-3^{\gamma)}$ & $-$ & $ \uparrow$  \\ \hline
14 & $1$ & $(0,0)$ & $0$ & $({\mathbf{E}})_{88}=-3^{\gamma)}$ & $-$ & $ \uparrow$\\ \hline \hline
15 & $1$ & $(0,0)$ & $|d| \in [3,15]$ & $0$ & $ - $ & $+$
\\ \hline \hline
16 & $1$ & $(\pm c,0)$, $c=5$ & $0$ & $0$ & $-$ & $+$ \\ \hline
\hline
17 & $1$ & $(0,0)$ & $\pm 10$ & $-3$ & $ - $ & $+$  \\ \hline
18 & $1$ & $(0,0)$ & $10$ & $1$ & $-$ & $\downarrow$  \\ \hline
19 & $1$ & $(0,0)$ & $5 \cdot \underline{\underline{x}}^{\delta)}$ & $1$
& $-$ & $+ $ \\ \hline \hline
\multicolumn{7}{l}{}\\[-2ex]
\multicolumn{7}{l}{$\alpha$) ${\mathbf{C}} = (C_{1},C_{2}) \in {\mathbf{R}}^{7,14}$, where $C_{1}, C_{2} \in {\mathbf{R}}^{7,7}$.}\\
\multicolumn{7}{l}{$\beta$) $1_{\lambda^{2}}$ denotes a diagonal
  matrix with $1 + \lambda^{2}_{i}$ at $(i,i)$.}\\
\multicolumn{7}{l}{$\gamma$) all other elements vanish.}\\
\multicolumn{7}{l}{$\delta$) $\underline{\underline{x}}$ denotes a
  diagonal matrix with space depending diagonal elements.}
\end{tabular}
\caption{\label{tabelle1}Results of the numerical experiments for
  various $\lambda$-systems. A number in the column parameter denotes a
  diagonal matrix with the number on the diagonal. For the notation used in
  the column norm, please refer to the text.}
\end{center}
\end{minipage}
\end{center}
\end{table}
The numbers in the parameter columns denote the value of the diagonal
elements of the corresponding diagonal matrix.
In the cases in which we studied a range of diagonal elements, we
condense by combining to parameter intervals with similar behaviour.
The observed development of the constraint norm $||\Delta_{{\cal
    C}}||$ and the pseudodifference norm $||\Delta_{{\cal P}}||$ is
described using the following notation:
\begin{description}
\item[$-/+$:] The norm is smaller/greater than that of the unextended
  system in the whole domain of time integration.
\item[$\uparrow/\downarrow$:] The norm is smaller/greater than that of the
  unextended system after small integration times and greater/smaller at
  the end of the integration.
\end{description}
The parameter choices for the run~1 were based on our expectations
explained in section~\ref{ZWEI}.
With this choice we were able to reduce the growth of the violation
of constraints, but we neither made the constraint submanifold an
attractor nor did we improve the pseudodifference at the end of the evolution.
Since we observed that the variables $\lambda$ grew faster and to
larger values than we had expected, we added terms proportional to
$\lambda^2$ (run~2--4), to increase the damping for non-vanishing $\lambda$s.
We did not observe any significant change in the behaviour.
This may have two reasons: Either the additional damping is
too weak, a change in the parameters $\mathbf{D}$ and $\mathbf{E}$
alone is not sufficient, or these parameters are not the
appropriate slots.
To obtain a better understanding of the effect of the single parameters 
we performed the numerical experiments~5--19 whose results we are going 
to discuss in the following subsections.
\subsection{Influence of the parameter B}
In the experiments~5--7, we studied the influence of the 
parameter $\mathbf{B}$ setting $\mathbf{B}$ proportional to the
unit matrix, $\mathbf{B}= b \ \underline{\underline{1}}$, and varying $b$.
The matrix ${\mathbf{E}}$ equals the unit matrix in all cases.
The choice of the parameter $b$ changes some of the characteristics of the
$\lambda$-system.
Absolute values of $b$ greater than $1$ imply that the $\lambda$-system
has characteristic speeds larger than the speed of light.
To avoid any influence of the grid boundary treatment on the
numerical solution in the physical part of the grid --- in the case $|b| >1$
the outer grid boundary is no longer causally disconnected from the
physical part of the grid --- we moved the grid boundaries further out,
something one can easily afford to do in a 1D calculation with
moderate values $b$.
For our purposes this was sufficient and we, therefore, did not
discretise the analytic treatment of the initial boundary value
problem for the $\lambda$-system which would guarantee that no
constraint violation is fed in from the grid boundary.
For $|b|>1$ the maximally allowed time step is smaller than that of the
unextended system due to the Courant-Friedrichs-Lewy condition.
To compare numerical experiments for systems with different values $b$,
we use the same, most restrictive time step for all runs.

\begin{figure}[htpb]
\noindent
\begin{minipage}[t]{.46\linewidth}
 \centering\epsfig{figure=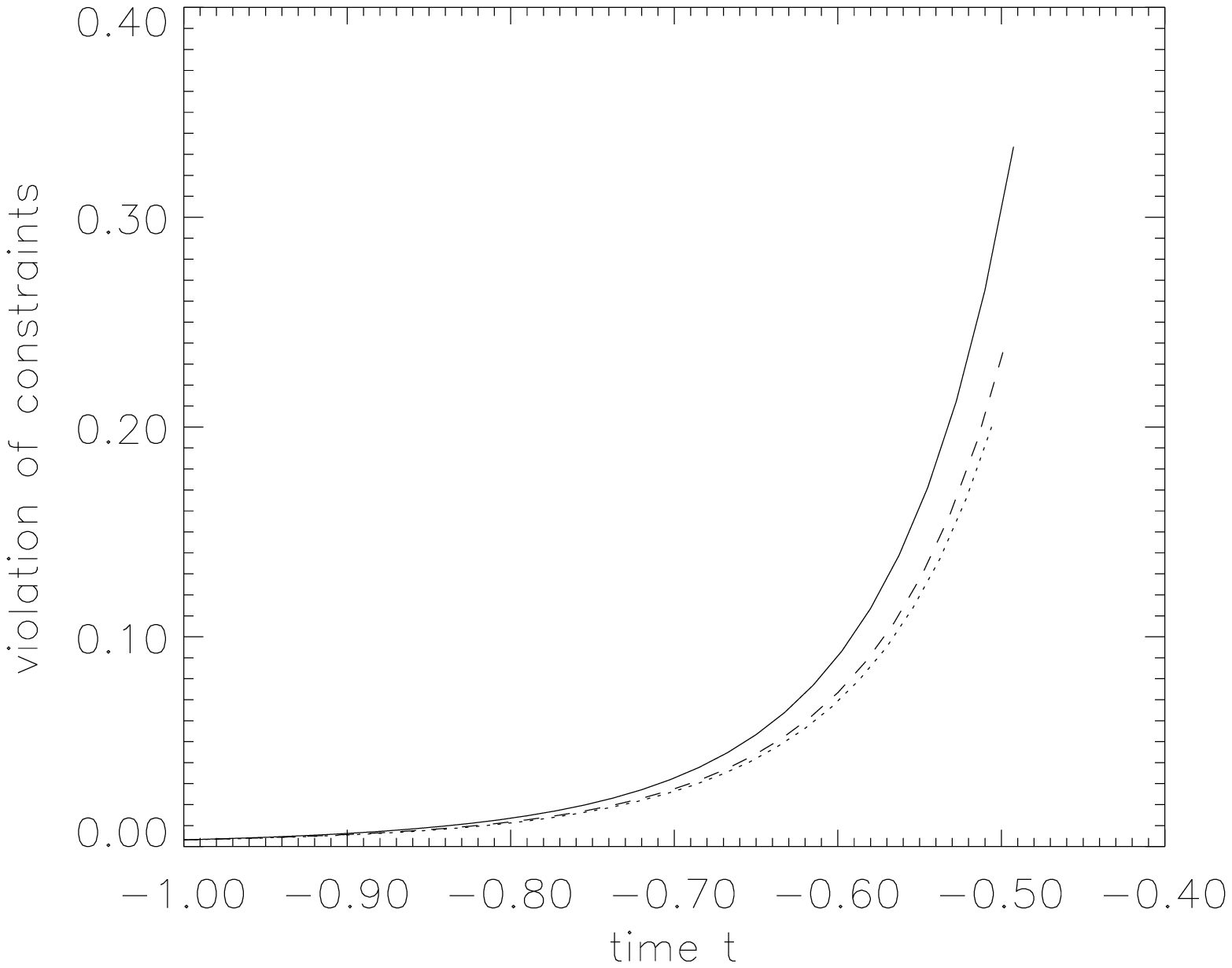,width=\linewidth}
 \caption{Constraint norm $||\Delta_{{\cal C}}||(t)$
   for the runs 5 with $b=-1$ (dashed line) and 6 (dotted line) in
   comparison to the unextended system (solid line).}
 \label{plot4}
\end{minipage}\hfill
\begin{minipage}[t]{.46\linewidth}
 \centering\epsfig{figure=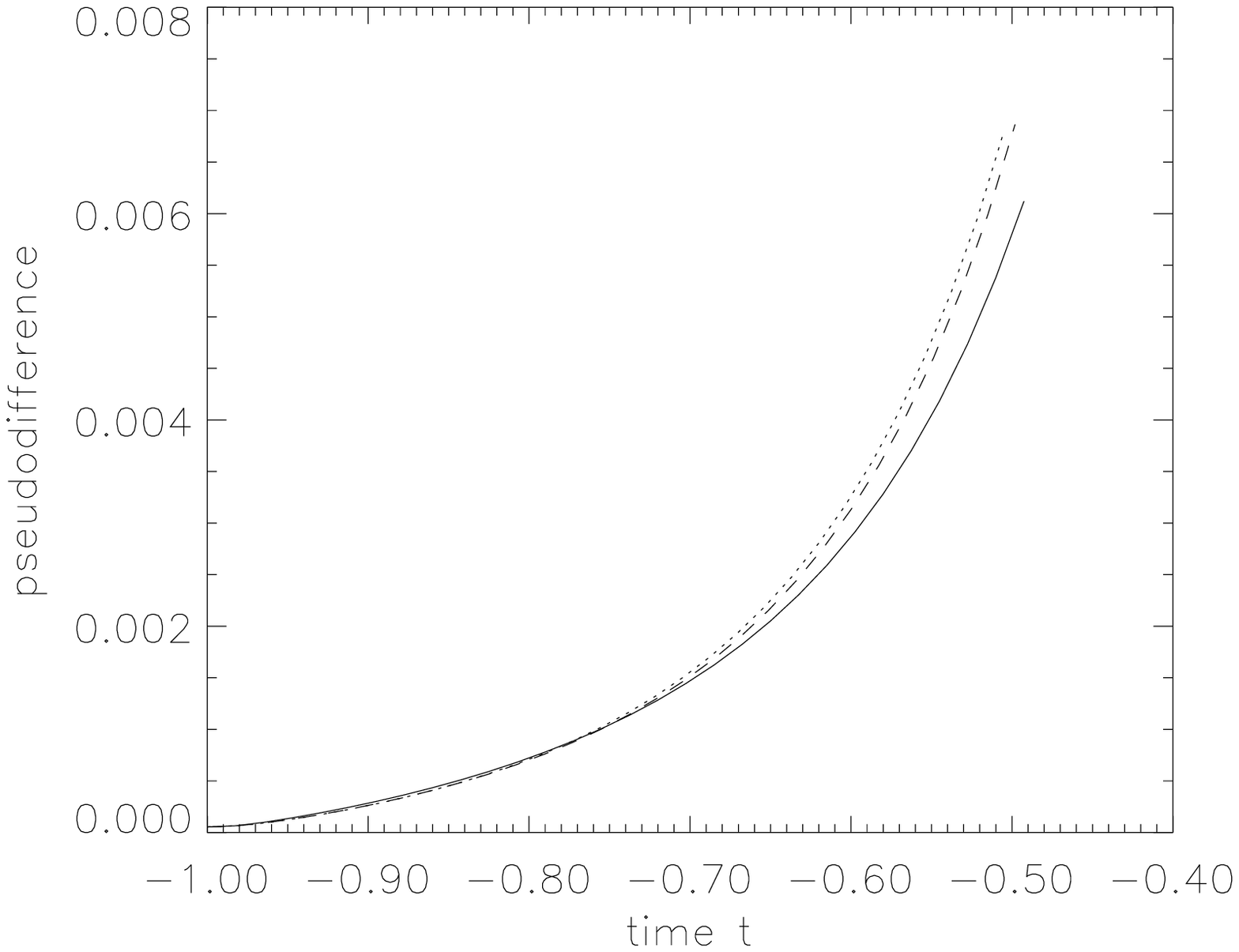,width=\linewidth}
 \caption{Pseudodifference norm $||\Delta_{{\cal P}}||(t)$ for the
   runs 5 with $b=-1$ (dashed line) and 6 (dotted line) in comparison
   to the unextended system (solid line).}
 \label{plot5}
\end{minipage}
\end{figure}

For small values of $b$ we do not observe a significant change in the
behaviour of the $\lambda$-system compared to the unextended system, as can be seen in the
FIG.~\ref{plot4} and \ref{plot5}.
In these figures we plot the constraint norm~(\ref{normconstr}) and the
pseudodifference norm~(\ref{normpseudo}) for $b=-1$ (dashed line) and
for $b=1.2$ (dotted line) in addition to the unextended system (solid
line).
The curves are very similar, as are all other curves from the runs
summarised as run number~5 in TABLE~\ref{tabelle1}.
There is a slightly reduced growth in the constraint norm and an
increase in the pseudodifference norm at the end of the evolution.

For a value of $b$ significantly greater than $1$ (run~7 and
FIG.~\ref{plot2} and \ref{plot3}) the constraint norm
decreases initially and is always smaller than in the unextended
system.
In contrast, as already mentioned, the pseudodifference norm is
always larger than in the unextended system.
\subsection{Influence of the parameter $\mathbf{E}$}
\begin{figure}[htpb]
\noindent
\begin{minipage}[t]{.46\linewidth}
 \centering\epsfig{figure=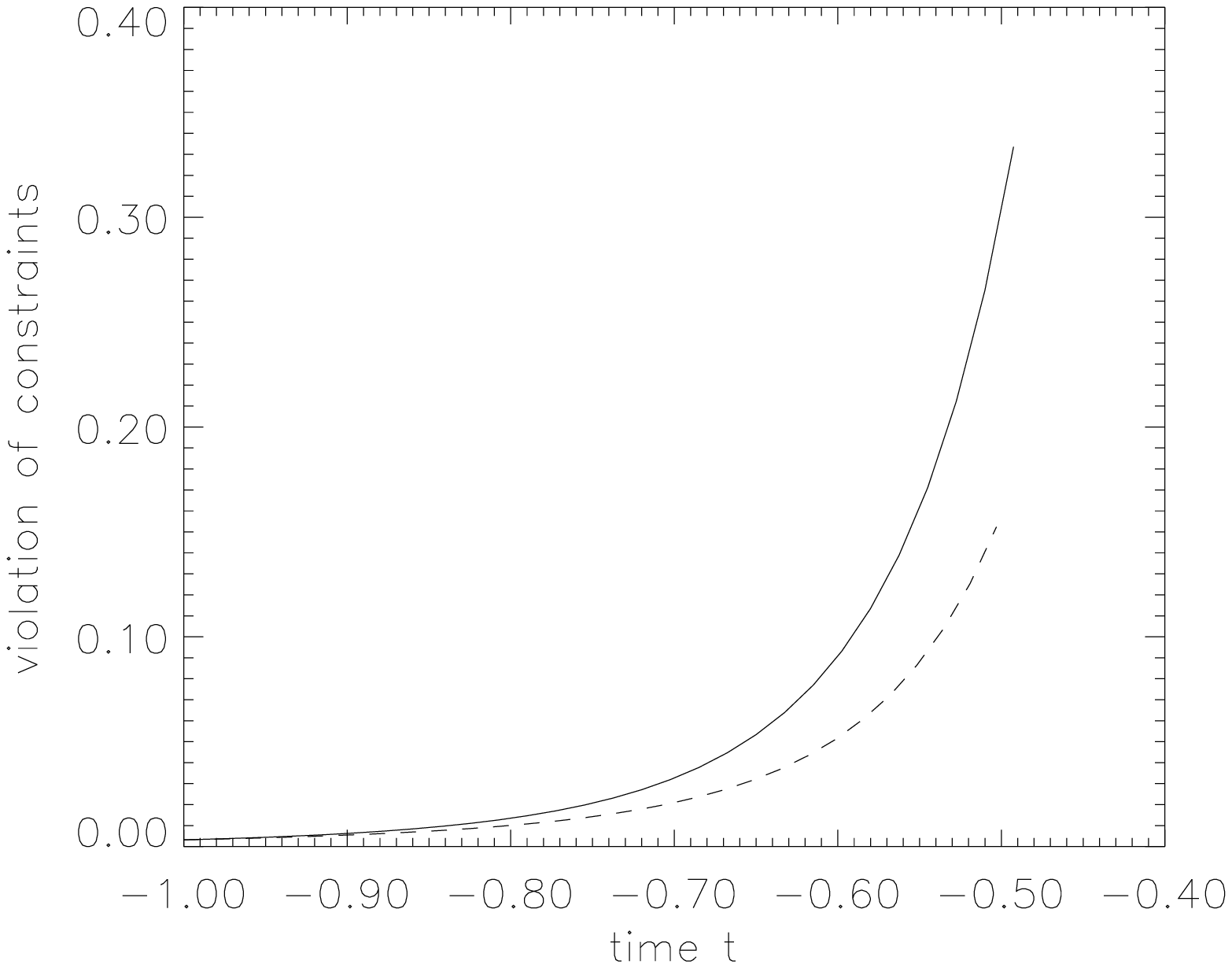,width=\linewidth}
  \caption{Constraint norm $||\Delta_{{\cal C}}||(t)$ for run 10 in
    comparison to the unextended system (solid line).}
 \label{plot6a}
\end{minipage}\hfill
\begin{minipage}[t]{.46\linewidth}
 \centering\epsfig{figure=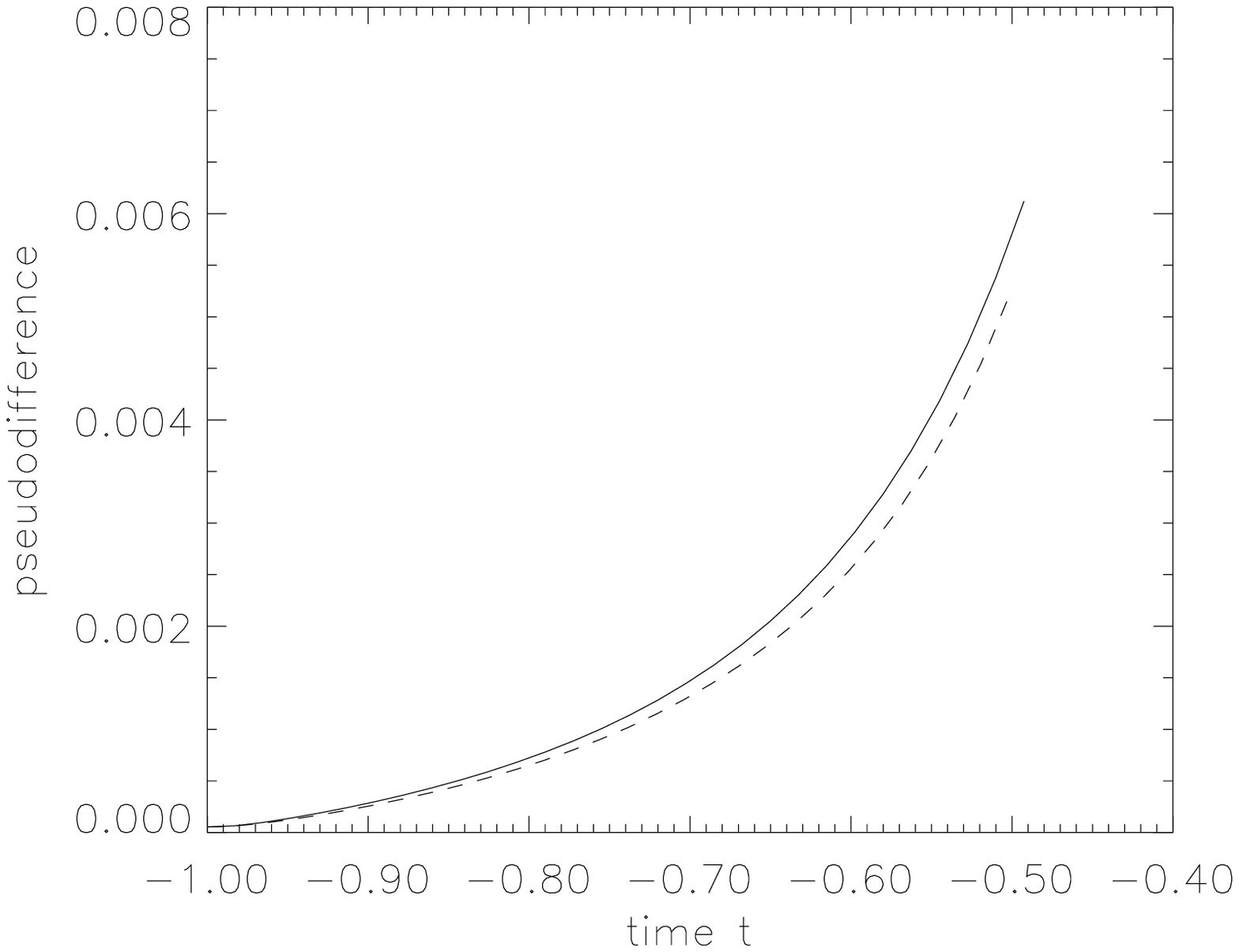,width=\linewidth}
 \caption{Pseudodifference norm $||\Delta_{{\cal P}}||(t)$ for run 10
   in comparison to the unextended system (solid line).}
 \label{plot7a}
\end{minipage}
\end{figure}

In runs~8-14 we studied the role of the parameter $\mathbf{E}$, fixing
the value of ${{\mathbf{B}}}$ to ${{\mathbf{B}}}=
\underline{\underline{1}}$. A non-zero value of ${{\mathbf{B}}}$ is
indeed necessary if $\lambda$ is to measure the violation of the
constraints.
We choose the value ${{\mathbf{B}}}= \underline{\underline{1}}$, as the
characteristic speeds determined by $\mathbf{B}$ then agree with speed
of light which seems to be a natural choice for the field equations of
general relativity.
In addition, the results in the previous subsection were rather insensitive to
the exact value of the parameter ${{\mathbf{B}}}$ in this part of the
parameter space.
Changing the value of diagonal elements for a constant diagonal matrix
$\mathbf{E}$ in experiments~8-11, we found the best results for the violation
of the constraints and the pseudodifference for a value of ${\mathbf{E}} =
-3 \ \underline{\underline{1}}$ (FIG.~\ref{plot6a} and FIG.~\ref{plot7a}),
where the variables $\lambda$ are amplified by the ${\mathbf{E}} \lambda$-term
in the $\lambda$-system. 
With this choice of parameters, the pseudodifference norm can
be slightly improved during the whole numerical integration up to $t = -0.5$, 
but we stress that this improvement is not significant.
We checked the results for this run after a longer integration time
(up to $t=-0.3$). We then found a worse pseudodifference norm compared to the
unextended system.  
In runs~12-14 we put only single diagonal elements of
$\mathbf{E}$ to -3, $({\mathbf{E}})_{11}$ and $({\mathbf{E}})_{88}$,
which influence directly those two constraints which are most
vehemently violated. 
These constraints are the constraints for the quantities $h_{11}$ and
$E_{11}$. 
Only affecting both constraints with our choice of parameters can
improve the numerical evolution of the constraints and the
pseudodifference compared to the unextended system.
However, in comparison to run~10, the results are worse.  
\subsection{Influence of the parameter $\mathbf{D}$}
\begin{figure}[htpb]
\noindent
\begin{minipage}[t]{.46\linewidth}
 \centering\epsfig{figure=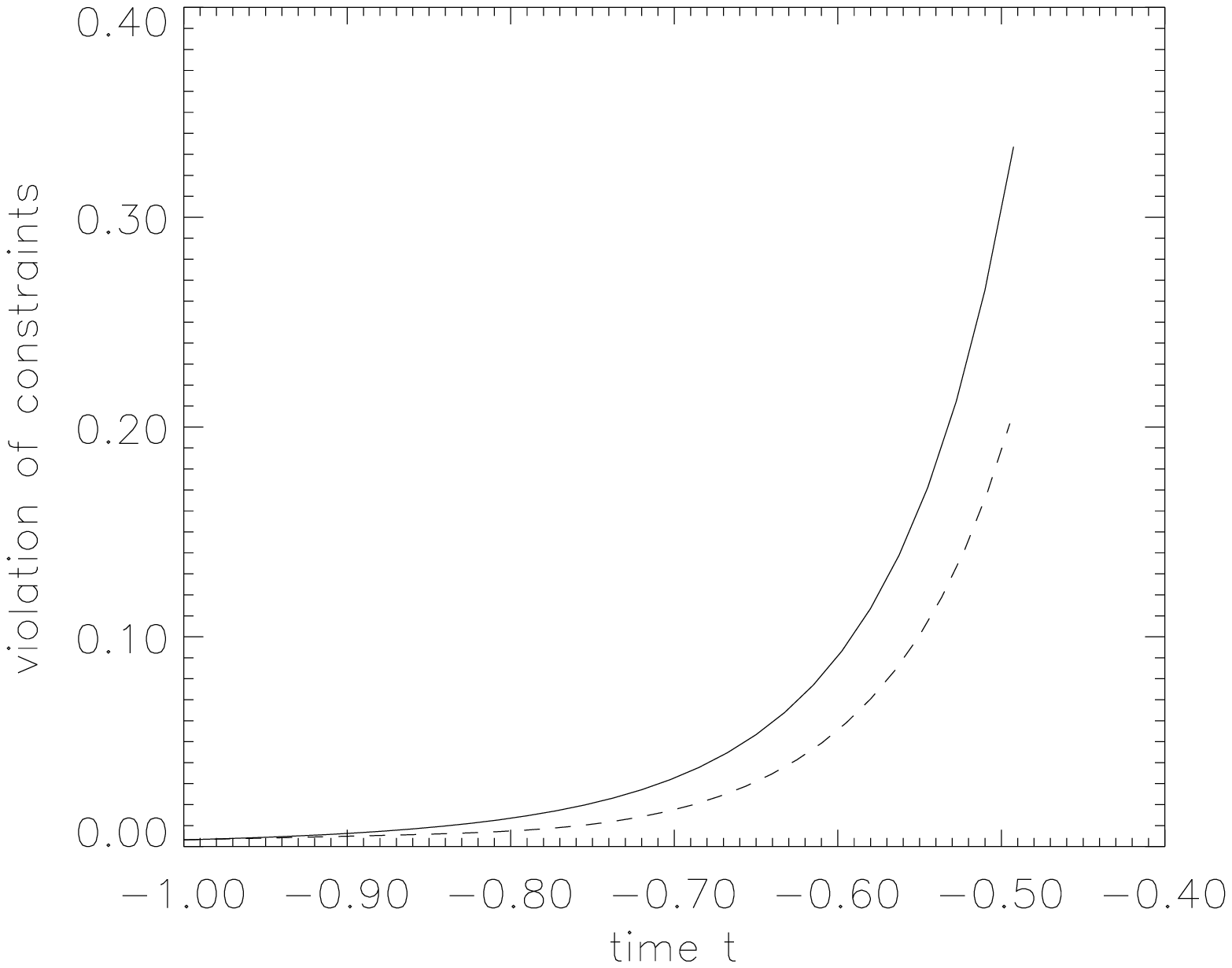,width=\linewidth}
  \caption{Constraint norm $||\Delta_{{\cal C}}||(t)$ for run~15 
    with $d=10$ in comparison to the unextended system (solid line).}
 \label{plot6}
\end{minipage}\hfill
\begin{minipage}[t]{.46\linewidth}
 \centering\epsfig{figure=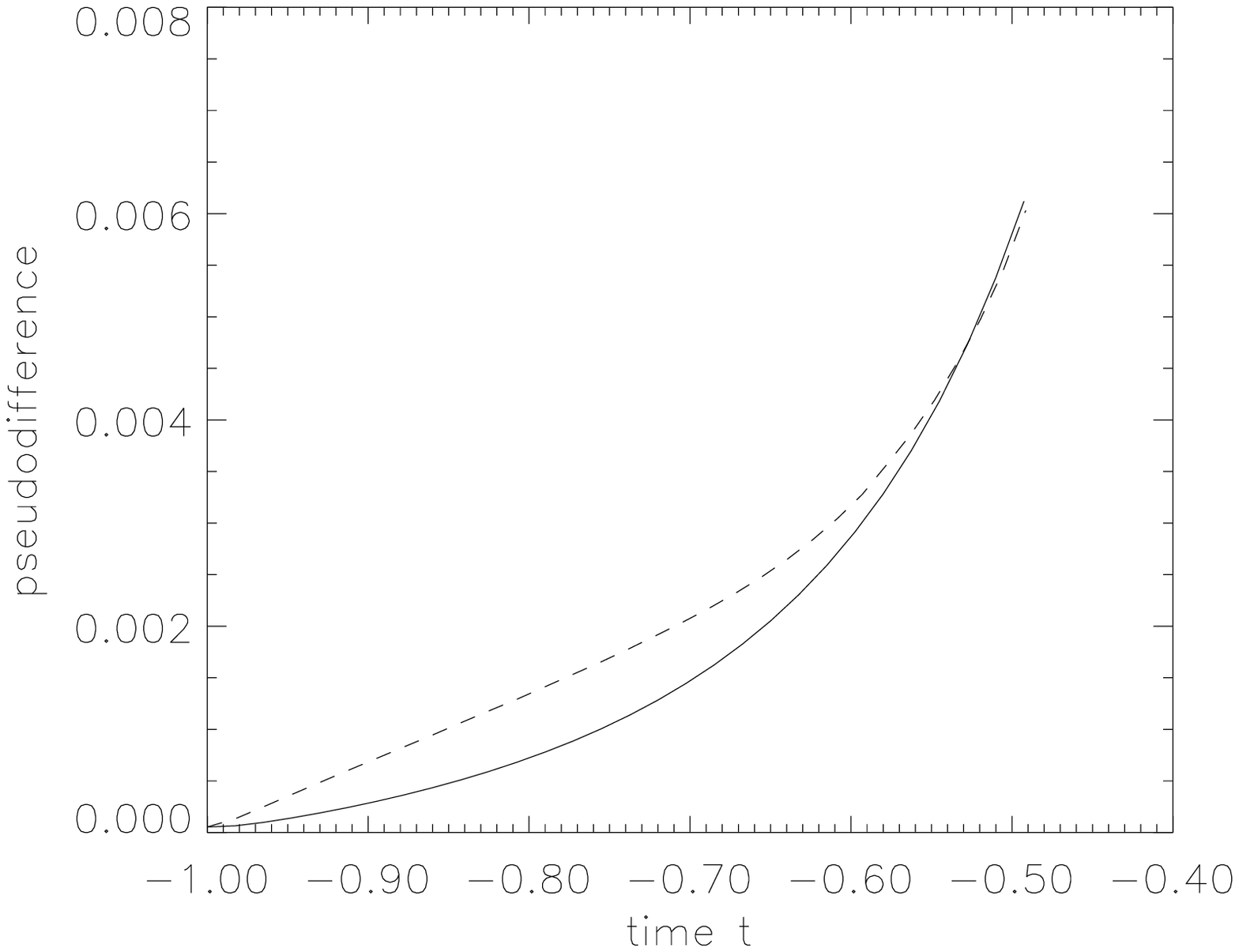,width=\linewidth}
 \caption{Pseudodifference norm $||\Delta_{{\cal P}}||(t)$ for
   run~19 in comparison to the unextended system (solid line).}
 \label{plot7}
\end{minipage}
\end{figure}

With the experiment~15, we studied the influence of a constant,
diagonal matrix $\mathbf{D}$ together with a matrix ${\mathbf{B}} =
\underline{\underline{1}}$.
In all runs with non-vanishing $\mathbf{D}$ the violation of
constraints could be improved (see e.g. FIG.~\ref{plot6} for $d=10$)
at the cost of a worse pseudodifference. 
The constraint and the pseudodifference norm are identical for runs with
${\mathbf{D}}=d \ \underline{\underline{1}}$ and ${\mathbf{D}}=-d \
\underline{\underline{1}}$. 
This property results from a symmetry of all time evolution variables
in~(\ref{lambda}) under the simultaneous transition $\mathbf{D} \to
-\mathbf{D}$, $x \to -x$ for this specific choice of parameters. 
For the A3-solution, all evolution variables at a fixed time are
even/odd functions on the space coordinate $x$. 
For a diagonal, constant matrix $\mathbf{D}$, the term
$\mathbf{D} \lambda$ couples to an even/odd function $f_i$ the
corresponding $\lambda_i$, which has, according to~(\ref{lambda3}),
the opposite symmetry, as it measures the corresponding constraint,
which involves one space derivative of $f_i$.
\subsection{Influence of the parameter $\mathbf{C}$}
In the experiments with a non-vanishing matrix $\mathbf{C}$ (run~16) the
violation of the constraints was only improved at the cost of a worse
pseudodifference. 
As the result for the constraint norm is similar to that in
FIG.~\ref{plot6}, and the result for the pseudodifference norm is
qualitatively given by FIG.~\ref{plot3}, we refrain from presenting
figures for run~16.
\subsection{Influence of the parameters $\mathbf{D}$ and $\mathbf{E}$}
Motivated by the good results for the pseudodifference in run~10 and for the
constraints in run~15,
we studied the correspondence of non-vanishing, diagonal
parameters $\mathbf{D}$ and $\mathbf{E}$ for a parameter ${\mathbf{B}}
= \underline{\underline{1}}$ in runs~17-19. 
Using the same parameters $\mathbf{B}, \mathbf{C}$ and $\mathbf{E}$ as
in run~10, but now with a non-vanishing $\mathbf{D}$, we could not
improve the numerical solution in run~17. 
Hence, in runs~18-19, we again stuck to our original choice of
$\mathbf{E}=\underline{\underline{1}}$. In run~18, the
pseudodifference first increased, before approaching the curve of the
unextended system at integration times of about $t = -0.5$ (FIG.~\ref{plot7}).
In experiment~19, we chose the matrix $\mathbf{D} = 5x \
\underline{\underline{1}}$ in order to keep the symmetries of the A3
solution.
Again, the constraints were only improved at the cost of a worse solution. 
\section{Conclusion and Outlook}
\label{SECHS}
The freedom in extending a system of evolution equations with
constraints to a $\lambda$-system is huge.
For the conformal field equations of general relativity, we have explored
the effect of what we thought are natural choices in the $\lambda$-system, 
analysing the influence on the quality of the new system's numerical solution.
We found that we were able to significantly reduce the violation of
the constraints, but we also found that this improvement did not imply
a smaller numerical error.

Furthermore, we found that the significant improvement in the violation
of the constraint did not prevent the solution from eventually running away
from the constraint submanifold, i.e.~the $\lambda$-systems used do
not inherit the property of asymptotic stability from the linear
system.
Similar experience with the semi-linear SU(2)-Yang-Mills
equations~\cite{BH} and the study of a simplified model
system~\cite{Sie} suggest that a more balanced choice of parameters of
the $\lambda$-system is needed to achieve an attractive force towards
the constraint submanifold for all times.
Recent analytic results by Heinz-Otto Kreiss and Peter H\"ubner give
sufficient conditions for asymptotic stability.
As we found a strong correlation  
between a smaller violation of the constraints and a worse numerical 
solution in the $\lambda$-system we suspect, however, that asymptotic 
stability does not necessarily imply a smaller numerical error.
\section*{Acknowledgement}
We would like to thank B. Schmidt and O. Brodbeck for helpful
discussions.
Peter H\"ubner would also like to thank the Albert-Einstein-Institut in
Golm for supporting the writing of the code described in~\cite{Hue}
and most of the work described in this paper by providing a postdoc
position. 
Florian Siebel thanks the members of the Albert-Einstein-Institut for the 
support during his Diplomarbeit.


\begin{thebibliography}{24}
\addcontentsline{toc}{chapter}{\numberline{}Literaturverzeichnis}

\bibitem{BFHR}
Brodbeck O., Frittelli S., H\"ubner P., Reula O., {\it J. Math. Phys.} {\bfseries 40} (1999) 909-923.

\bibitem{Cho}
Choptuik M.W., {\it Phys. Rev. D} {\bfseries 44} (1991) 3124-3135.

\bibitem{Det}
Detweiler S., {\it Phys. Rev. D} {\bfseries 35} (1987) 1095-1099.

\bibitem{Fri}
Friedrich H., {\it Proc. R. Soc. London A} {\bfseries 375} (1981) 169-184.

\bibitem{Fri1}
Friedrich H., {\it Proc. R. Soc. London A} {\bfseries 378} (1981) 401-421.

\bibitem{Fri2}
Friedrich H., {\it Commun. Math. Phys.} {\bfseries 91} (1983) 445-472.

\bibitem{Hue}
H\"ubner P., {\it Class. Quantum Grav.} {\bfseries 16} (1999) 2145-2164.

\bibitem{Hue2}
H\"ubner P., {\it Class. Quantum Grav.} {\bfseries 15} (1998) L21-L25.

\bibitem{Hue3}
H\"ubner P., {\it Class. Quantum Grav.} {\bfseries 16} (1999) 2823-2843.

\bibitem{LWe}
Lax P.D., Wendroff B., {\it Comm. Pure Appl. Math}. {\bfseries 17} (1964) 381.

\bibitem{BH}
Brodbeck O., H\"ubner P., unpublished work.

\bibitem{Sie}
Siebel F., {\it Diploma thesis} LMU Munich (1999).

\end{thebibliography}
\end{document}